\documentstyle[prb,aps,twocolumn,floats,color,graphicx]{revtex}

\newcommand{\xc}{\mbox{$\psi$}}
\newcommand{\xcp}{\mbox{$\psi_p$}}
\newcommand{\xcs}{\mbox{$\psi_s$}}
\newcommand{\xci}{\mbox{$\psi_i$}}

\newcommand{\cv}{\mbox{$\phi$}}
\newcommand{\cvp}{\mbox{$\phi_p$}}
\newcommand{\cvs}{\mbox{$\phi_s$}}
\newcommand{\cvi}{\mbox{$\phi_i$}}

\newcommand{\oxc}{\mbox{$\omega_x$}}

\newcommand{\op}{\mbox{$\omega_p$}}
\newcommand{\os}{\mbox{$\omega_s$}}
\newcommand{\oi}{\mbox{$\omega_i$}}

\newcommand{\opo}{\mbox{$\omega_p^0$}}
\newcommand{\oso}{\mbox{$\omega_s^0$}}
\newcommand{\oio}{\mbox{$\omega_i^0$}}

\newcommand{\gx}{\mbox{$\gamma_x$}}
\newcommand{\gc}{\mbox{$\gamma_c$}}
\newcommand{\gp}{\mbox{$\gamma_p$}}
\newcommand{\gs}{\mbox{$\gamma_s$}}
\newcommand{\gi}{\mbox{$\gamma_i$}}

\newcommand{\fp}{\mbox{$F_p$}}
\newcommand{\fs}{\mbox{$F_s$}}
\newcommand{\fbp}{\mbox{$f_p$}}
\newcommand{\fbs}{\mbox{$f_s$}}

\newcommand{\xa}{\mbox{$\mbox{\scriptsize{\it X}}$}}
\newcommand{\xap}{\mbox{$\mbox{\scriptsize{\it X}}_p$}}
\newcommand{\xas}{\mbox{$\mbox{\scriptsize{\it X}}_s$}}

\newcommand{\ca}{\mbox{$\mbox{\scriptsize{\it C}}$}}
\renewcommand{\cap}{\mbox{$\mbox{\scriptsize{\it C}}_p$}}
\newcommand{\cas}{\mbox{$\mbox{\scriptsize{\it C}}_s$}}

\newcommand{\xf}{\mbox{$|\mbox{\scriptsize{\it X}}|$}}
\newcommand{\xfp}{\mbox{$|\mbox{\scriptsize{\it X}}_p|$}}
\newcommand{\xfs}{\mbox{$|\mbox{\scriptsize{\it X}}_s|$}}
\newcommand{\xfi}{\mbox{$|\mbox{\scriptsize{\it X}}_i|$}}

\newcommand{\cfp}{\mbox{$|\mbox{\scriptsize{\it C}}_p|$}}

\newcommand{\ok}{\mbox{$\omega_c({k})$}}
\newcommand{\okp}{\mbox{$\omega_c({k_p})$}}
\newcommand{\oks}{\mbox{$\omega_c({k_s})$}}
\newcommand{\oki}{\mbox{$\omega_c({k_i})$}}

\newcommand{\kp}{\mbox{${{\bf k}_p}$}}
\newcommand{\ks}{\mbox{${{\bf k}_s}$}}
\newcommand{\ki}{\mbox{${{\bf k}_i}$}}

\newcommand{\kbar}{\mbox{$\overline{\kappa}$}}

\newcommand{\half}{\mbox{$\frac{1}{2}$}}
\newcommand{\twelfth}{\mbox{$\frac{1}{12}$}}
\newcommand{\degree}{\mbox{$^{\circ}$}}

\begin{document}

\title{Parametric Processes in a Strong-Coupling Planar Microcavity}

\author{D. M. Whittaker}

\address{21, Bermuda Road, Cambridge, CB4 3JX, United Kingdom.}

\date{\today}

\maketitle

\begin{abstract} 
I present a theoretical treatment of parametric scattering in strong coupling
semiconductor microcavities to model experiments in which
parametric oscillator behaviour has been observed. The model
consists of a non-linear excitonic oscillator coupled to a cavity mode
which is driven by the external fields, and predicts the output power,
below threshold gain and spectral blue shifts of the parametric
oscillator. The predictions are found to be in excellent agreement
with the experimental data
\end{abstract}

Recent experimental studies have demonstrated that very large optical
non-linearities can be obtained in resonantly pumped strong-coupling
microcavities.  The excitations of these structures are polaritons,
mixed modes which are part exciton, part cavity photon, and the
non-linearity is due to interactions between the exciton components,
which cause the polaritons to scatter off each other. This leads to a
parametric process where a pair of pump polaritons scatter into
non-degenerate signal and idler modes, while conserving energy and
momentum. The scattering is particularly strong in microcavities
because the unusual shape of the dispersion, shown in the inset to
Fig.\ref{pumpangle}, makes it possible for pump, signal and idler
all to be on resonance at the same time.

A further important property of planar microcavities is the
correspondence between the in-plane momentum of each polariton mode
and the direction of the external photon to which it couples.
This makes it quite straight forward to investigate parametric
scattering using measurements at different angles to access the various
modes. Two types of process have been studied in this way:
parametric amplification, where the scattering is stimulated by
excitation of the signal mode with a weak probe field, and parametric
oscillation, where there is no probe and a coherent population in the
signal mode appears spontaneously.

Parametric amplification in microcavities was first observed by
Savvidis {\it et al}\/\cite{savvidis2000}, using ultrafast pump-probe
measurements.  The structure was pumped on the lower polariton branch
at an incident angle of 16.5\degree. Narrow band gains of up to 70
were observed in the region of the polariton feature for a probe at
0\degree, along with idler emission at 35\degree. These are a set of
angles for which the pair scattering resonance condition is satisfied.
A related scattering process has been studied by Huang {\it et
al}\/\cite{huang2000} using two pump beams at $\pm45\degree$ and a
probe at normal incidence.  The experimental results of
Ref.\onlinecite{savvidis2000} have been modelled by Ciuti {\it et
al}\/\cite{ciuti2000a} using a microscopic quantum treatment of
polariton-polariton interactions.

Parametric oscillator behaviour has been observed Stevenson {\it et
al}\/\cite{stevenson2000} and Baumberg {\it et
al}\/\cite{baumberg2000}, in CW experiments with the pump incident on
the lower polariton branch at the `magic' angle of about
16\degree. Above a threshold pump intensity, strong signal and idler
beams were observed at about 0\degree\/ and 35\degree, without any
probe stimulation. The coherence of these beams was demonstrated by
significant spectral narrowing, proving that they are due to a
parametric process rather than resonantly enhanced incoherent
photoluminescence.  Houdr\'e {\it et al}\/\cite{houdre2000}, have also
observed a nonlinear emission at 0\degree\/ for a structure pumped at
10\degree. However, in this experiment the pair scattering resonance
condition is not satisfied, suggesting that different physics may
be involved.

The purpose of this paper is to develop a simple classical model
which provides a unified treatment of both amplifier and oscillator in
the CW regime. This model is based on the textbook treatment of
parametric phenomena in systems such as LiNbO$_3$\/\cite{shen}. Indeed, the
microcavity behaviour has similar characteristics to a typical
doubly-resonant parametric oscillator, where just the signal and idler
modes are cavity resonances -- the pump resonance is mainly important
in enhancing the strength of the nonlinear effects. However, there are
also significant novel aspects to the model: the
microcavity operates in the strong coupling regime, where the modes
are cavity polaritons not simple photons, and instead of a
non-resonant $\chi^{(2)}$ nonlinearity, the exciton provides a
highly resonant $\chi^{(3)}$ effect.

\vspace{-1em}

\section{Model}

\vspace{-0.5em}

The theoretical model is a classical treatment of a non-dispersive
exciton mode, $\xc(r)$, with energy \oxc, coupled to a cavity mode,
$\cv(r)$, with dispersion \ok, which is driven by the external
fields. To account for broadening processes, \oxc\/ and \ok\/ are
taken to be complex energies, with imaginary parts \gx\/ and \gc\/
respectively. The exciton mode is non-linear, with potential energy
$V(\xc)=\half \oxc^2 \xc^2 + \twelfth \kappa \xc^4$. 

To model the parametric processes, the cavity is driven by harmonic
plane waves, consisting of a pump with amplitude \fp\/ at $(\op,\kp)$
and a probe with amplitude \fs\/ at (\os,\ks).  The cavity and exciton
modes are also expressed as a sum of plane waves at $(\op,\kp)$ and
(\os,\ks), plus an idler at $(\oi=2\op-\os,\ki=2\kp-\ks)$. The cavity
mode is linear, so the equations of motion for the pump, signal and
idler mode separate out, giving
\begin{eqnarray}
\label{cavityterms}
(\okp^2-\op^2) \cvp + g \xcp &=& \fp
\\ \nonumber
(\oks^2-\os^2) \cvs + g \xcs &=& \fs
\\ \nonumber
(\oki^2-\oi^2) \cvi + g \xci &=& 0,
\end{eqnarray}
where $g$ is the strength of the coupling between the exciton and the
cavity photon. The exciton equations are more complicated because the
non-linearity generates many terms at different frequencies and
wave-vectors.  Only the terms at frequencies \op, \os, \oi\/ are
retained here: the others are at very different frequencies, such as $3 \,
\op$, or are weak, less than $O(\xcp^2)$. This leaves
\begin{eqnarray}
\label{excitonterms} 
(\oxc^2-\op^2) \xcp + g \cvp + 
\kappa |\xcp|^2 \xcp + 2 \kappa \xcs \xci \xcp^* &=& 0 
\\ \nonumber
(\oxc^2-\os^2) \xcs + g \cvs + 
2 \kappa |\xcp|^2 \xcs +  \kappa \xcp^2 \xci^* &=& 0 
\\ \nonumber
(\oxc^2-\oi^2) \xci + g \cvi + 
2 \kappa |\xcp|^2 \xci +  \kappa \xcp^2 \xcs^* &=& 0
\end{eqnarray}

These equations can be simplified by using Eqs.(\ref{cavityterms}) to
eliminate the cavity photon fields \cv\/ and write everything in terms
of the exciton fields \xc. It is also convenient to approximate
$\oxc^2-\op^2 \approx 2 \oxc (\oxc-\op)$ etc and define
$\Omega=g/\oxc$, $\kbar=\half\kappa/\oxc$ and $f= \half F /
\oxc$. Then Eqs.(\ref{excitonterms}) become

\makebox[2.5in][r]{
\parbox{3.3in}{
\begin{small}
\begin{mathletters}
\label{model} 
\begin{eqnarray}
\left(\oxc+ \kbar |\xcp|^2 -\op-\frac{(\Omega/2)^2}{\okp-\op}\right) 
\xcp + 2\,\kbar \xcs \xci \xcp^* &=& \frac{-\Omega/2}{\okp-\op} \, \fbp 
\label{model1} \\ 
\left(\oxc+ 2 \kbar |\xcp|^2 -\os-\frac{(\Omega/2)^2}{\oks-\os}\right) 
\xcs + \kbar \xcp^2 \xci^* &=& \frac{-\Omega/2}{\oks-\os} \, \fbs 
\label{model2} \\ 
\left(\oxc + 2 \kbar |\xcp|^2 -\oi-\frac{(\Omega/2)^2}{\oki-\oi}\right) 
\xci + \kbar \xcp^2 \xcs^* &=& 0
\label{model3} 
\end{eqnarray}
\end{mathletters}
\end{small}
}}

Eqs.(\ref{model}) constitute the basic model for parametric
processes in a microcavity. The terms in $|\xcp|^2$ represent the
renormalisation of the exciton energy due to the pump population.
The other non-linear terms provide the scattering, which is
the main interest here: $\kbar \xcp^2 \xci^*$ and $\kbar \xcp^2
\xcs^*$ in (\ref{model2},\ref{model3}) describe the
build up of the population in the signal and idler modes, while
$2\,\kbar \xcs \xci \xcp^*$ in (\ref{model1}) represents the
corresponding pump depletion.

It is often useful to make the simplification of considering a
situation where the pump, signal and idler energies are all close to
the corresponding polariton resonance values, and the broadenings are
small compared to the Rabi splitting $\Omega$.  Then it is a good
approximation to replace the polariton response by a single Lorentzian
function at each ${\bf k}$, with strength $\xf^2$, where \xa\/ is the
exciton amplitude (Hopfield coefficient) for the mode. The driving
terms on the right hand side of Eqs.(\ref{model}) can similarly be
approximated by $(\ca/\xa) f$, where \ca\/ is the photon
amplitude. Then, Eqs.(\ref{model}) reduce to
\begin{mathletters}
\label{simplified}
\begin{eqnarray}
\frac{1}{\xfp^2}(\opo+i\gp-\op) \, \xcp + 
2 \,\kbar \xcs \xci \xcp^* &=& \frac{\cap}{\xap} \, \fbp 
\label{simplified1} \\ 
\frac{1}{\xfs^2}(\oso+i\gs-\os) \, \xcs + \kbar \xcp^2 \xci^* &=& 
\frac{\cas}{\xas} \, \fbs 
\label{simplified2} \\ 
\frac{1}{\xfi^2}(\oio+i\gi-\oi) \, \xci + \kbar \xcp^2 \xcs^* &=& 0
\label{simplified3}
\end{eqnarray} 
\end{mathletters}
where $\opo$, $\oso$ and $\oio$ are the polariton resonance
frequencies, and \gp, \gs, \gi\/ the corresponding widths. In writing
the equations in this form, the exciton renormalisation is effectively
ignored, though it can be considered to be included as a
renormalisation of the polariton frequencies: at each point on the
dispersion, it leads to a blue shift of approximately $2 \kbar \xf^2
|\xcp|^2$.  For low pump powers, where the blue shift is small
compared to the pump polariton width, and when $\op=\opo$, \xcp\/ can
be approximated, using Eq.(\ref{simplified1}), by $\xcp \approx -i
\cap \xap^* \fbp / \gp$, and the blue shift is
\begin{equation}
\label{blueshift}
\delta \omega^0
\approx 2 \kbar \xf^2 \, \frac{\cfp^2 \xfp^2}{\gp^2} \, I_p
\end{equation}
where $I_p=|\fbp|^2$ is the pump intensity.

It is interesting to compare the present classical model with the
treatment in Ref.\onlinecite{ciuti2000a}, which gives a good fit to
the pump-probe parametric amplifier experiments of
Ref.\onlinecite{savvidis2000}. This treatment was based on a
quantum mechanical picture of the exciton-exciton scattering
process. However, with the approximations that were made, 
Eqs(1-3) of Ref.\onlinecite{ciuti2000a}, contain essentially
the same physics as Eqs.(\ref{simplified}) here. Of course, using a
microscopic model gives a value for the non-linearity
$\kbar$. However, $\kbar$ only imposes a scale on the problem:
rescaling all the fields so $\xc \rightarrow
\xc/\sqrt{\kbar}$, $f \rightarrow f/\sqrt{\kbar}$,  effectively makes 
$\kbar=1$. 

\vspace{-1em}

\section{Parametric Amplifier}
\label{sec-amp}

\vspace{-0.5em}

In the parametric amplifier, both \fbp\/ and \fbs\/ are non-zero. It
is also helpful to assume that $\fbs \ll \fbp$, so that \xcs\/ and 
\xci\/ are small, and the pump depletion term in
Eqs.(\ref{model1},\ref{simplified1}) can be neglected.  Consider first
the situation when the probe, idler and pump satisfy the triple
resonance condition, so Eqs.(\ref{simplified}) can be used with
$\op=\opo$, $\os=\oso$ and $\oi=\oio$. Without the pump depletion
term, these equations are solved by eliminating \xcp\/ and \xci\/
using (\ref{simplified1}) and then (\ref{simplified3}), to get
\begin{equation}
\xcs = \frac{-i \, \cas \xas^* \fbs / \gs}
{1 - \kbar^2 \frac{\xfi^2 \xfs^2}{\gi \gs} 
\frac{\cfp^4 \xfp^4 }{\gp^4} \, |\fbp|^4}
\end{equation}
Dividing by the value of \xcs\/ without the pump, ie with
$\fbp=0$, gives the {\em internal gain} for the probe:
\begin{equation}
\label{gain}
\alpha_s = \frac{1}{1 - I_p^2/I_0^2}
\end{equation}
where
\begin{equation}
\label{threshold}
I_0 = \frac{\gp^2 \, \sqrt{\gs \gi}}{\kbar \, \cfp^2 \xfp^2 \xfs \xfi}
\end{equation}
The gain increases from unity at $I_p=0$ to become singular at
$I_p=I_0$. This suggests that $I_0$ represents the threshold pump
intensity for oscillation, which will indeed be
shown to be the case in the next section.

The previous discussion was limited to the case where all the fields
are on resonance. If this condition is not satisfied, it is still
possible to obtain an analytic solution when the exciton
renormalisation is neglected. Solving Eqs.(\ref{model}) with the
same weak probe approximation gives
\begin{equation}
\label{spectral}
\xcs=
\frac{-(\Omega/2) \fbs / \Lambda_s}
{1 - \kbar^2 
\frac{(\oks-\os)}{\Lambda_s} \frac{(\oki^*-\oi)}{\Lambda_i^*} 
\frac{(\Omega/2)^4}{|\Lambda_p|^4} 
\, |\fbp|^4 }
\end{equation}
where $\Lambda_p=(\oxc-\op)(\okp-\op)-(\Omega/2)^2$, and $\Lambda_s,
\Lambda_i$ are similarly defined. 

\begin{figure}[t]

\begin{center}
\vspace{-0.4in}
\mbox{
\includegraphics[scale=0.4]{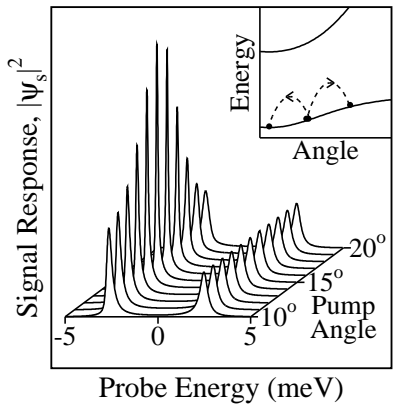}
}
\end{center}

\caption{ Dependence of the signal response on the pump angle, with a
probe at normal incidence. The pump amplitude $\fbp=0.292$, which
corresponds to $I_p=0.75 I_0$ for this geometry. Structure parameters
are: Rabi splitting $\Omega=5.0$ meV, zero detuning, exciton width
\gx = 0.25 meV, cavity width \gc = 0.25 meV, nonlinearity $\kbar=1$.
The inset shows the polariton dispersion with the pair scattering from 
pump to signal and idler modes.
}
\label{pumpangle}
\end{figure}

\begin{figure}[t]
\vspace{-0.3in}
\begin{center}
\mbox{
\includegraphics[scale=0.95]{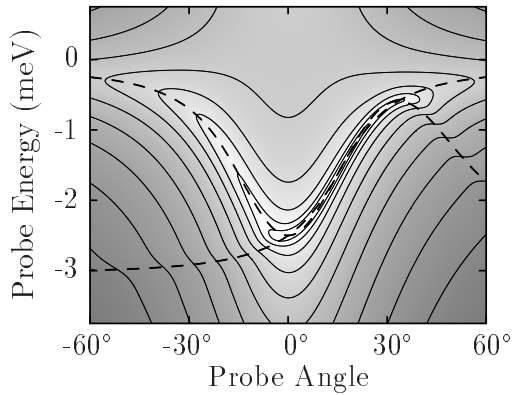}
}
\end{center}
\vspace{0.0in}
\caption{Contour plot of the signal response $|\xcs|^2$ as a function of probe
energy and incidence angle. The contours are logarithmically spaced,
and the light regions of background shading indicate the highest
intensities. Dashed lines show the single resonance conditions for
signal and idler.  The pump is on resonance at 16\degree,
and other parameters are as in
Fig.\ref{pumpangle}.}
\label{probemap}
\end{figure}

Fig.\ref{pumpangle} shows the signal response at normal incidence,
calculated using Eq.(\ref{spectral}), for different pump angles.  The
pump energy is varied to be on the polariton resonance for each
angle. The spectra show the two polariton features at normal
incidence, with clear gain on the lower branch for pump angles in the
region of 16\degree: for this pump angle, the pair scattering
resonance condition is satisfied when the probe angle is 0\degree.

In Fig.\ref{probemap}, the signal response is mapped out as a function
of energy and angle with the pump kept on resonance at
16\degree. There is an enhanced probe response when either the signal
or the idler is on resonance -- these single resonance energies are
indicated by the dashed lines on the figure. The response is much
stronger for the signal resonance, because the probe couples to the
signal directly, but to the idler only via parametric
scattering. The strongest response occurs at the double resonances,
where the dashed lines intersect (at 0\degree, 16\degree\/ and
33.5\degree), but there is a long segment of the dispersion very close
to double resonance, where the gain remains high.

\vspace{-1em}

\section{Parametric Oscillator}
\label{sec-opo}

\vspace{-0.5em}

In the parametric oscillator regime, there is no probe, so $\fbs=0$,
but solutions can still be found with finite signal and idler fields.
Again, it is simplest to start with the triply resonant case. Focusing
on Eqs.(\ref{simplified2},\ref{simplified3}), taking the complex
conjugate of one of them, and treating \xcp\/ as a parameter, there
are non-zero solutions for
\xcs, \xci\/ only if the determinant of the coefficients is zero, 
that is:
\begin{equation}
\label{balance}
\kbar \, |\xcp|^2 \xfs \xfi  = \sqrt{\gs \gi}
\end{equation}
The physical interpretation of this condition is obvious -- for a
steady state solution, the generation rate of polaritons in the signal
and idler directions, on the left hand side, must equal the
(geometric) mean of the loss rates, $\sqrt{\gs \gi}$. This is only
possible with the value of \xcp\/ in Eq.(\ref{balance}). For a given
external driving field \fbp, the required value of \xcp\/ is attained
by the depletion of the pump polariton field due to the stimulated
scattering term.

The resulting signal intensity is calculated by using
Eq.(\ref{simplified2}) to write \xci\/ in terms of \xcs, then
substituting in (\ref{simplified1}) to obtain
\begin{equation}
|\xcs|^2 = \frac{\gi}{2 \kbar^2 \, |\xcp|^3 \xfi^2}
\frac{\cfp}{\xfp} \left( |\fbp| - \frac{|\xcp|}{\cfp \xfp} \gp \right)
\end{equation}
where $|\xcp|$ is now just a constant given by Eq.(\ref{balance}).
Since the emitted signal intensity, $I_s$, is proportional to
$|\xcs|^2$, this relationship is of the form
\begin{equation}
\label{power}
I_s \propto \sqrt{I_p} - \sqrt{I_0},
\end{equation}
where $I_0$ is the same threshold intensity as in Eq.(\ref{threshold}).
Of course $I_s \ge 0$, so this solution only exists when $I_p \ge 
I_0$.

The treatment can be extended to the case where the signal direction
is such that the signal and idler are not both on resonance, that is
the mismatch $\Delta=2\op-\oso-\oio \neq 0$.  The steady state
condition, Eq.({\ref{balance}) becomes
\begin{equation}
\label{unbalance}
\kbar \, |\xcp|^2 \xfs \xfi = 
[(\oso + i\gs - \os)  (\oio -i\gi - \oi)]^{\half} ,
\end{equation}
with, once again, $\oi=2 \op - \os$. A solution is only possible if
the right hand side of Eq.(\ref{unbalance}) is real, which requires
$\os-\oso=\Delta \gs/(\gs + \gi)$, and correspondingly
$\oi-\oio=\Delta \gi/(\gs + \gi)$.  The physical significance of this
requirement can be seen by looking at \xcs\/ and \xci: for the allowed
value of \os, $\gs |\xcs|^2/\xfs^2 = \gi |\xci|^2/\xfi^2$, that is,
the loss rates through the signal and idler modes are
identical. This is the Manley-Rowe relation for the 
parametric oscillator.

Continuing the solution for the signal intensity, for resonant pumping
the effect of the finite mismatch $\Delta$ is to shift the threshold,
so
\begin{equation}
\label{fullthreshold}
I_0(\Delta)=I_0(0)
\left( 1 + \frac{\Delta^2}{(\gs+\gi)^2} \right)^{\half}, 
\end{equation}
where $I_0(0)$ is the value given in Eq.(\ref{threshold}).  The
threshold is lowest when $\Delta=0$ and pump, signal and idler are all
on resonance.

These results show that the signal intensity is determined by the pump
depletion, which produces the value of $|\xcp|^2$ required by
Eqs.({\ref{balance},\ref{unbalance}). For $I_p>I_0$, $|\xcp|^2$ remains
unchanged, just as the population inversion in a conventional laser is
clamped at its threshold level.  Since the actual value of
$|\xcp|^2$ is unique to a particular pair of signal and idler
directions, in equilibrium there can only be one
finite signal amplitude.  It is easy to see what will happen in an out
of equilibrium situation when there is more than one signal. A signal
whose loss rate exceeds its generation rate will decay, while one for
which the generation rate exceeds the loss will grow.  So, in
a process akin to mode selection in a laser, the signal with the
lowest loss rate will dominate, depleting the pump until only it
survives.

\vspace{-1em}

\section{Discussion}

\vspace{-0.5em}

This model of the parametric oscillator makes two simple predictions
which can be checked against experiment: the $\sqrt{I_p}$ power
dependence in Eq.(\ref{power}), and the clamping of the pump polariton
amplitude, \xcp, to the value given in Eq.(\ref{balance}). The latter
effect should be observable as a saturation, above threshold, of the
blue shift of the polariton dispersion. Below the threshold the shift
is roughly linear with pump power, and it saturates at
the value given in Eq.(\ref{blueshift}) with $I_p=I_0$. For the signal
mode the saturated shift is thus $\delta \oso \approx 2 \sqrt{\gs
\gi} \, \xfs/\xfi$. Of course, there are other effects, not included
in the model, which can cause energy shifts. These include the exciton
renormalisation due to the signal field, and at higher powers the
break down of the strong coupling regime. The former effect can be
estimated and is small: the signal field will give an energy shift of
$\kbar \xfs^2 |\xcs|^2$, which is about 10\% of the saturated $\delta
\oso$ when $I_p=2I_0$, using the same structural parameters as in
Fig.\ref{pumpangle}.

\begin{figure}[t]

\begin{center}
\vspace{-0.3in}
\mbox{
\includegraphics[scale=0.35]{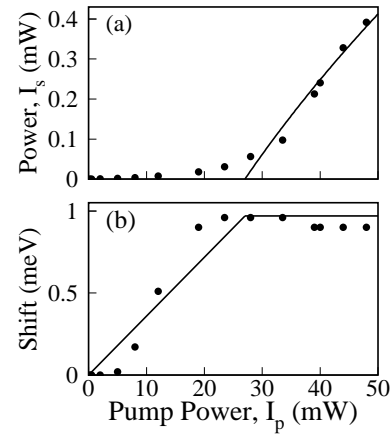}
}
\end{center}

\caption{
Theoretical fits to the experimental data (points) for (a) signal power and (b)
blue shift from Ref.\protect \onlinecite{baumberg2000}.
}
\label{experiment}
\end{figure}

Fig.\ref{experiment} shows how these predictions compare with the
experimental results in Ref.\onlinecite{baumberg2000}. The signal
power is fitted to the form $I_s \propto \sqrt{I_p} - \sqrt{I_0}$ of
Eq.(\ref{power}). Although a good fit is obtained, it is difficult to
distinguish the $\sqrt{I_p}$ behaviour from a simple linear increase
with the limited range of data above threshold. The support for the
model provided by the blue shift data is rather better: the
experiments show a clear saturation of the blue shift, and the saturation
shift of just under 1 meV agrees very well with the prediction of
$\delta \oso \approx 0.97$meV obtained using the experimental widths
$\gs=0.57$meV and $\gi=0.80$meV.

The predicted form of the gain in Eq.(\ref{gain}) is not found in the
pump-probe experiments of Ref.\onlinecite{savvidis2000}, where the
gain increases exponentially with $I_p$. However, the gain mechanism
in the ultrafast measurements differs from the CW parametric
amplification discussed here. The peak pump powers in the experiments are
orders of magnitude greater than the threshold $I_0$, but oscillator
behaviour is not seen because there is insufficient time for the
signal to build up just from the incoherent
photoluminescence. However, the probe pulse seeds the signal, starting
an an exponential growth which lasts as long as the pump.  This gives
the observed exponential dependence on $I_p$. Although the CW gain
below threshold predicted here has not been seen experimentally, it
should be observable in structures which show parametric oscillator
behaviour.

To conclude, it has been shown that the parametric oscillator
behaviour recently observed in microcavities can be well explained by
a simple classical treatment of a non-linear exciton strongly coupled
to a cavity mode. The main non-classical effects that are missing from
this model are the incoherent photoluminescence, discussed
theoretically in a recent paper by Ciuti {\it et
al.},\cite{ciuti2000b} and quantum statistical effects in the
oscillator behaviour around the threshold. It is also possible that at
higher excitation powers the signal intensity will become large enough
for the appearance of new physics due to polariton-polariton
interactions in the coherent state.

I gratefully acknowledge the contributions made by P. B. Littlewood,
M. S. Skolnick, J. J. Baumberg, P. G. Saviddis and R. M. Stevenson in
discussions about this work.

\end{document}